\def\al{\alpha}
\def\be{\beta}

\def\de{\delta}

\def\th{\theta}

\def\ta{\tau}

\def\fr#1#2{\frac{#1}{#2}}

\def\ket#1{|{#1}\rangle}

\def\lsim{\mathrel{\rlap{\lower4pt\hbox{\hskip1pt$\sim$}}
    \raise1pt\hbox{$<$}}}
\def\gsim{\mathrel{\rlap{\lower4pt\hbox{\hskip1pt$\sim$}}
    \raise1pt\hbox{$>$}}}

\def\etal {{\it et al.}}

\newcommand{\beq}{\begin{eqnarray}}
\newcommand{\eeq}{\end{eqnarray}}
\def\to{\rightarrow}

\def\no{\nonumber}

\def\maxf{10}
\def\minf{10}

\documentclass{ws-procs9x6-cpt16}
\begin{document}

\newcommand{\refeq}[1]{(\ref{#1})}
\def\etal {{\it et al.}}

\title{
Test of Lorentz Violation with Astrophysical Neutrino Flavor
}

\author{
Teppei Katori$^1$ , 
Carlos~A.~Arg\"{u}elles$^{2,3,4}$,
and Jordi Salvado$^{2,3,5}$\\
}

\address{
$^1$School of Physics and Astronomy, Queen Mary University of London, E1 4NS, UK \\
$^2$Department of Physics, University of Wisconsin, Madison, WI 53706, USA \\
$^3$Wisconsin IceCube Particle Astrophysics Center, Madison, WI 53706, USA \\
$^4$Physics Department, Massachusetts Institute of Technology, Cambridge, MA 02139, USA \\
$^5$Instituto de F\'{i}sica Corpuscular (IFIC), CSIC-Universitat de Val\`{e}ncia, E-46071 Valencia, Spain
 }

\begin{abstract}
The high-energy astrophysical neutrinos recently discovered by IceCube
opened a new way to test Lorentz and CPT violation through
the astrophysical neutrino mixing properties. 
The flavor ratio of astrophysical neutrinos is a very powerful tool 
to investigate tiny effects caused by Lorentz and CPT violation~\cite{fratio}.
There are 3 main findings;
(1) current limits on Lorentz and CPT violation in neutrino sector are not tight and they allow for any flavor ratios,
(2) however, the observable flavor ratio on the Earth is tied with the flavor ratio at production,
this means we can test both the presence of new physics
and the astrophysical neutrino production mechanism simultaneously,
and (3) the astrophysical neutrino flavor ratio is one of the most stringent tests of Lorentz and CPT violation.


\end{abstract}

\bodymatter

\section{Neutrino mixing}
The propagation of neutrinos are eigenstates of Hamiltonian, however,
the production and detection of neutrinos are flavor eigenstates.
This is the source of ``neutrino oscillations'', which is the topic of
the 2015 Nobel Prize of Physics~\cite{SuperK,SNO},
and the 2016 Breakthrough Prize
in Fundamental Physics~\cite{SuperK,SNO,K2K,T2K,DayaBay,KamLAND}. 

\vspace{5mm}
The flavor eigenstates $\ket{\nu_\alpha}$ is written by the superposition of
the propagation eigenstates $\ket{\nu_i}$ with a unitary matrix $V(E)$, which
diagonalize the Hamiltonian in the flavor basis, 
\beq
  \ket{\nu_\alpha}= \sum_i V_{\alpha i}(E) \ket{\nu_i}~,~
  H(E)= V(E)^{\dagger}
\left(\begin{array}{ccc}
\Delta_1(E)  & 0 & 0 \\
0 &\Delta_2(E) & 0 \\ 
0 & 0 & \Delta_3(E)
\end{array}\right)
V(E),\no
\label{eq:prophamiltonian}
\eeq
where $\Delta_i(E)$ is the $i$th eigenvalue. 
Note through this article we assume there are only 3 generations.
Also we consider a simple case where the Hamiltonian only depends on the neutrino energy. 
Given the Hamiltonian and solving the evolution of an initial flavor
state $\ket{\nu_\alpha}$ over a distance $L$,
the probability of measuring a flavor state $\ket{\nu_\beta}$ is obtained,  
\beq
P_{\nu_{\al} \to \nu_{\be}}(L,E)
&=&\de_{\al \be}
-4  \sum_{i>j}Re(V_{\al i}^{*}V_{\be i}V_{\al j}V_{\be j}^{*})
\sin^2\left(\fr{\Delta_i-\Delta_j}{2}L\right)+\cdots \no
\label{eq:osc}
\eeq
However, not all experiments actually measure the neutrino as it ``oscillates'',
most notably solar neutrinos.
When neutrinos propagate a significantly longer distance than their oscillation lengths, 
coherent behavior is washed out. 
Then, neutrinos do not oscillate, but ``mix'' incoherently. 
In this situation, Eq.~\ref{eq:osc}
can be written only with mixing matrix elements, 
\beq
\bar P_{\nu_{\al} \to \nu_{\be}}(E)&=&
\sum_{i}\left|V_{\al i}(E)\right|^2\left|V_{\be i}(E)\right|^2~.
\label{eq:mix}
\eeq
This is the mixing probability of
astrophysical neutrinos which propagate mostly in the vacuum. 
Any new physics in the vacuum,
such as Lorentz and CPT violation,
would induce anomalous neutrino mixings
which may be imprinted in the astrophysical neutrino flavor ratio measured on the Earth.
This is expected to be more sensitive than kinematic tests of
Lorentz and CPT violation with astrophysical neutrinos~\cite{Diaz_UHEnu}.

\section{Astrophysical Neutrino Flavor Ratio}
Details of this analysis are described in the published paper~\cite{fratio}. 
The astrophysical neutrino flux on the Earth's surface, $\phi^{\oplus}_\beta (E)$,
can be obtained by convoluting Eq.~\ref{eq:mix} with the astrophysical neutrino flux at the production,
$\phi^p_\alpha$, {\it i.e.}, $\phi^{\oplus}_\beta (E)=\sum_\alpha \bar P_{\nu_{\al} \to \nu_{\be}}(E) \phi^p_\alpha(E)$. 
Then, an energy averaged flavor composition $\bar \phi^{\oplus}_\beta$ is obtained 
by integrating $E^{-2}$ power law production flux
within $\Delta E=$[$\maxf~\mathrm{TeV}$, $\minf~\mathrm{PeV}$] with 50 bins in $log_{10}$. 
Since the absolute flux of astrophysical neutrinos are not well known,
we focus on the relative flavor information, namely the flavor ratio, 
$\alpha^{\oplus}_\beta=\bar\phi^{\oplus}_\beta / \sum_\gamma \bar\phi^{\oplus}_\gamma$.

We use following Hamiltonian to look for new physics.
\beq
  H=\frac{1}{2E}U M^2 U^\dagger +\sum_n \left(\frac{E}{\Lambda_n}\right)^n \tilde U_n O_n \tilde U_n^\dagger
  = V^\dagger(E) \Delta V(E)~,
  \label{eq:hamiltonian}
\eeq
here, the first term is the standard neutrino mass matrix~\cite{NuFit},
and the rest are the effective operators of new physics.
For example, $n=0$ and $n=1$ terms can be interpreted
as the time component of the CPT-odd and CPT-even SME coefficients. 
This moment we do not take into account the spatial components,
mainly because of the lack of the spatial distribution information of astrophysical neutrinos. 

\begin{figure}
\begin{center}
\includegraphics[width=1.47in]{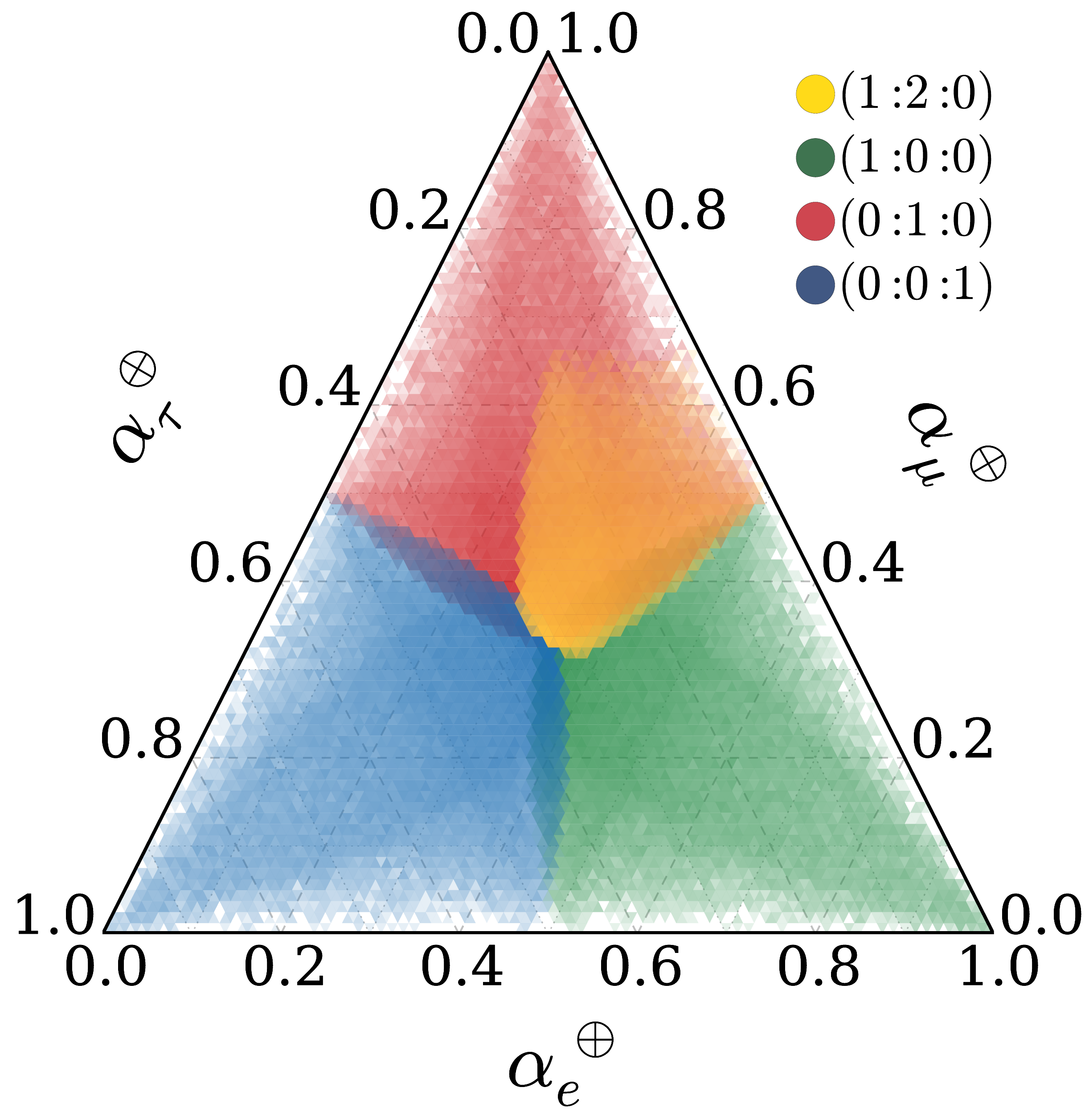}
\includegraphics[width=1.47in]{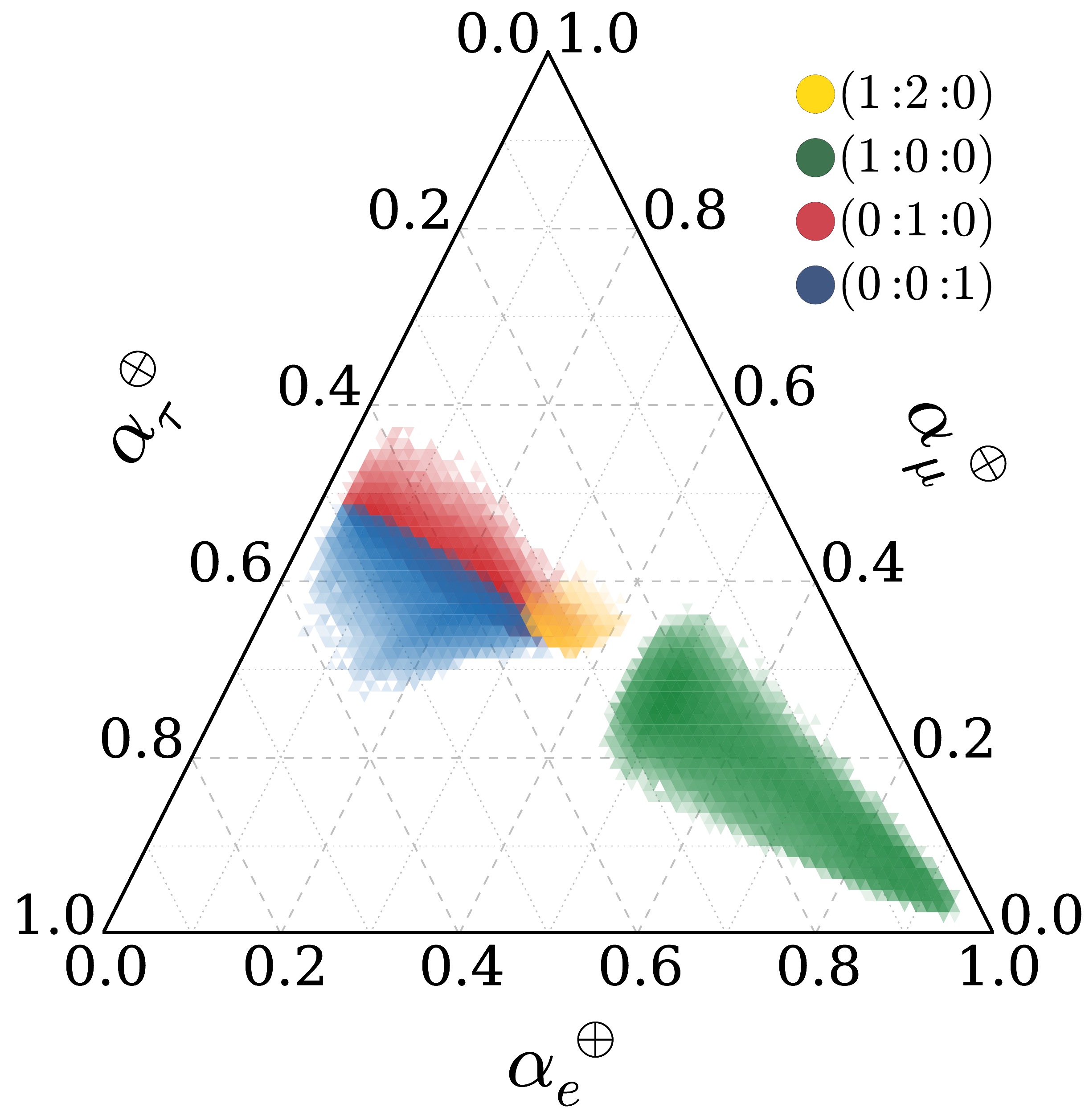}
\includegraphics[width=1.47in]{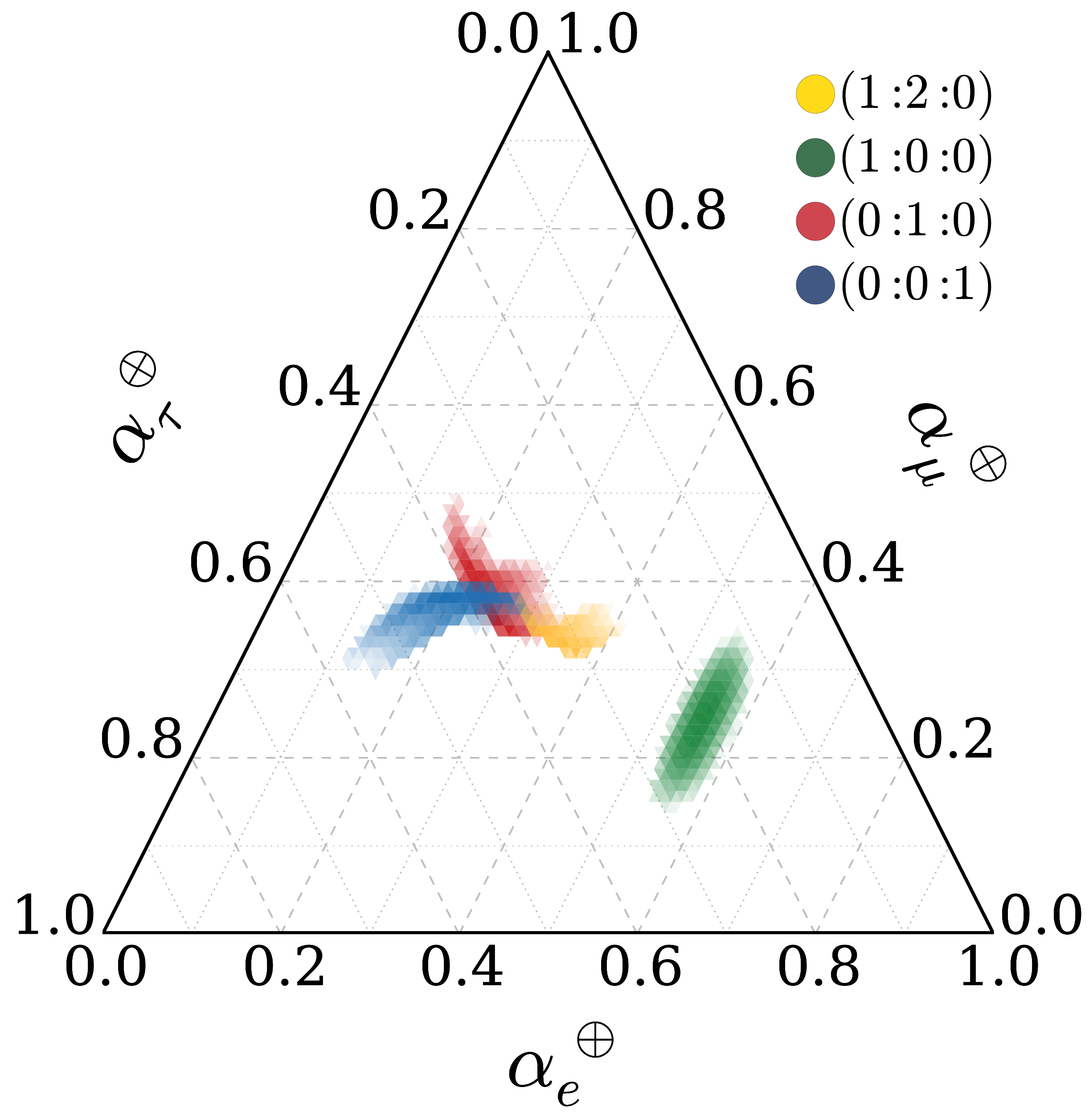}
\end{center}
\caption{
 Allowed flavor ratio region using for the CPT-even SME coefficients~\cite{fratio}.
 The left plot corresponds roughly to the current limits ($c= 1.0\times 10^{-27}$);
 the middle plot corresponds to $c=1.0\times 10^{-30}$,
 while the right plot corresponds to $c=3.2\times 10^{-34}$.
}
\label{fig:fig4}
\end{figure}

Figure~\ref{fig:fig4} shows our results for the $n=1$ case
(CPT-even Lorentz violation, or ``c'' coefficients in minimal SME).
The result is obtained by using the standard GSL library~\cite{GSL}
and the Armadillo C++ linear algebra library~\cite{armadillo}.
In our approach, we first set the ``scale'' of new physics, then we sample 100,000 points
of the mixing matrix $\tilde U_n$ by the anarchy sampling scheme~\cite{anarchy}.
Neutrino parameters are also sampled by choosing symmetric errors~\cite{NuFit}
from a Gaussian distribution, except $\de_{CP}$ which is sampled from a flat distribution.
The results are all from the lower octanct ($\th_{23}<\frac{\pi}{2}$) and with normal mass ordering,
however, these choices only give minor effects. 
Here we show the result with 3 scales,
the left plot of Fig.~\ref{fig:fig4} is $c= 1.0\times 10^{-27}$,
this is around the current best limit for CPT-even SME coefficients in the neutrino sector by
Super-K and IceCube atmospheric neutrino data~\cite{SuperK_LV,IceCube_LV}.
As you see, whole flavor triangle is covered,
this means by current limits of Lorentz violation we could expect any flavor ratio from observations.

Next, we set the scale to be $c=1.0\times 10^{-30}$ and $c=3.2\times 10^{-34}$, 
these are the scale of CPT-even Lorentz violation terms when
they become comparable to the neutrino mass term with $E=35$~TeV and $E=2$~PeV. 
The available phase space of flavor triangle shrinks,
this means now the size of new physics is getting smaller compared to the neutrino mass, 
and eventually the neutrino mass will dominate the flavor ratio.
In fact, the right plot is almost identical to the standard flavor ratio with
different assumptions on neutrino production models~\cite{fratio,Bustamante,Palladino}. 

There are four distinct regions, depending on different assumptions of
the astrophysical neutrino production mechanism.
However, regardless of the details of the production process,
the predicted flavor ratio is always around the center ($\nu_e:\nu_\mu:\nu_\ta\sim 1:1:1$),
and only by the presence of new physics,
will it deviate from the central area.
Therefore, the exotic flavor ratio may nail down both the presence of new physics,
and the production mechanism of astrophysical neutrinos.

Finding the flavor ratio is experimentally challenging~\cite{IceCube}, 
due especially to the separation of $\nu_e$ and $\nu_\ta$ largely relying on the simulation input~\cite{Mena}. 
The measurements can be improved with current detectors, 
but future IceCube-Gen2~\cite{Gen2} and KM3NeT~\cite{KM3NeT}
will have a better chance of studying 
the astrophysical neutrino flavors from their superior detectors. 

\section*{Acknowledgments}
TK thanks to the organizers of CPT 2016 for their hospitality
during my stay at Bloomington, Indiana, USA.
We also thank Shivesh Mandalia for the careful reading of this manuscript.  
This work is supported by Science and Technology Facilities Council, UK. 


\end{document}